\documentclass[10pt]{article}
\usepackage{graphicx}

\usepackage{revsymb}
\usepackage{moreverb}
\usepackage{dcolumn}    
\usepackage{bm}              
\usepackage{epstopdf}    
\usepackage{rotating}
\usepackage{float}
\usepackage{amsthm}
\usepackage{amssymb}
\usepackage{latexsym}
\usepackage{url}
\usepackage{color}
\usepackage{lineno,xcolor}
\usepackage{fixltx2e}

\def\Title#1{\begin{center} {\Large #1 } \end{center}}
\def\Author#1{\begin{center}{ \sc #1} \end{center}}
\def\Address#1{\begin{center}{ \it #1} \end{center}}

\newcommand\pubblock{\rightline{\begin{tabular}{l} Proceedings of the Second Annual LHCP\\ \pubnumber\\
         \pubdate  \end{tabular}}}

\newenvironment{Abstract}{\begin{quotation} \begin{center} 
             \large ABSTRACT \end{center}\bigskip 
      \begin{center}\begin{large}}{\end{large}\end{center} \end{quotation}}

\newenvironment{Presented}{\begin{quotation} \begin{center} 
             PRESENTED AT\end{center}\bigskip 
      \begin{center}\begin{large}}{\end{large}\end{center} \end{quotation}}

\def\Acknowledgements{\bigskip  \bigskip \begin{center} \begin{large}
             \bf ACKNOWLEDGEMENTS \end{large}\end{center}}

\def\beq{\begin{equation}}
\def\eeq#1{\label{#1}\end{equation}}
\def\eeqn{\end{equation}}

\def\beqa{\begin{eqnarray}}
\def\eeqa#1{\label{#1}\end{eqnarray}}
\def\eeqan{\end{eqnarray}}

\let\bar=\overbar

\def\Dslash{\not{\hbox{\kern-4pt $D$}}}
\def\dslash{\not{\hbox{\kern-2pt $\del$}}}

\def\msb{{\bar{\ssstyle M \kern -1pt S}}}

\usepackage{color}

 \newcommand\pubnumber{ }

\newcommand\pubdate{\today}

\def\affiliation{
On behalf of the ALICE Collaboration, \\
Department of Physics \\
University of Houston, Houston, TX 77204, U.S.A }

\begin{document}

\large
\begin{titlepage}
\pubblock

\vfill
\Title{ Strangeness production in two-particle azimuthal correlations on the near and away side measured with ALICE in pp collisions at $\sqrt{s}$=7 TeV}
\vfill

\Author{ Sandun Jayarathna}
\Address{\affiliation}
\vfill
\begin{Abstract}

Two-particle azimuthal correlations allow one to study high-$p_{\rm T}$ parton fragmentation without full jet reconstruction. Enhancements of the azimuthal correlations are seen at $\Delta \varphi \approx 0$ and $\Delta \varphi \approx \pi$,  resulting from back-to-back jet fragmentation in the parton center-of-mass system. We present the current status of the study of correlations between charged trigger particles and associated strange baryons ($\Lambda$) and mesons (K$_{\textsubscript{S}}^{0}$) in pp collisions at $\sqrt{s}$ = 7 TeV. A data-driven feeddown correction for $\Lambda$ is also presented, which could allow a more accurate calculation of the primary $\Lambda/$K$_{\textsubscript{S}}^{0}$ ratio in jets and the underlying event.

\end{Abstract}
\vfill

\begin{Presented}
The Second Annual Conference\\
 on Large Hadron Collider Physics \\
Columbia University, New York, U.S.A \\ 
June 2-7, 2014
\end{Presented}
\vfill
\end{titlepage}
\def\thefootnote{\fnsymbol{footnote}}
\setcounter{footnote}{0}
%

\normalsize 

\section{Introduction}

An enhancement of the strange baryon-to-meson ratio is observed in heavy-ion collisions for the intermediate $p_{\rm T}$ region with respect to pp collisions \cite{Abelev:2013xaa, Abelev:2007xp}.  This is typically explained in terms of collective flow or quark recombination \cite{Muller:2012zq}. Two-particle azimuthal correlations are a powerful tool to explore such particle production mechanisms. The yields in the near- and away-side peaks are usually associated with jet fragmentation, while the yields in the underlying event are usually associated with multiple partonic interactions and initial- and final-state radiation \cite{ALICE:2011ac}. The main aim of this ongoing analysis is to extract strange particle yields from correlations with an unidentified charged trigger particle in pp collisions at $\sqrt{s} = 7$ TeV. The goal is to eventually analyze the strange baryon and meson production in the peak and underlying regions in pp collisions in order to map out the various hadronization processes in elementary collisions. In these proceedings, we present the current status of the analyses performed with a high-$p_{\rm T}$ trigger particle and strange baryons ($\Lambda + \bar\Lambda $)  and mesons (K$^{0}_{\textsubscript{S}}$) as associated particles.
\section{Analysis Technique}
ALICE is a dedicated heavy-ion physics experiment at the LHC with the primary objective of studying properties of the Quark Gluon Plasma \cite{Carminati:2004fp}. The analysis is performed  on a sample of 260 $\times$ 10$^{6}$ minimum bias (MB) events from pp collisions at a center-of-mass energy of $\sqrt{s}$ = 7 TeV in the 2010 run of the LHC. Triggering is done using a hit in the Silicon Pixel Detector or a signal in the V0 detectors. In addition, we impose a primary vertex Z position cut of $|$z$|$$<$ 10 cm. The charged particles are reconstructed using tracking information from the Inner Tracking System and the TPC \cite{Carminati:2004fp}. The trigger particle is the leading particle in a single well-defined interval of 6 $ < p_{\rm T} <  $12 GeV/\textit{c}. Associated strange particles are reconstructed through their decay topology (K$_{\textsubscript{S}}^{0} \rightarrow \pi^{+} + \pi^{-},  \Lambda \rightarrow p + \pi^{-},  \bar\Lambda \rightarrow \bar {p} + \pi^{+}$). Since these neutral particles decay into a V-shaped topology, they are commonly known as V$^{0}$ particles. Topological cuts are placed on the V$^{0}$s to reduce the combinatorial background, and different topological cuts are used for  K$^{0}_{\textsubscript{S}}$ and $\Lambda$ due to their different decay kinematics \cite{Aamodt:2011zza}.

Once the topological selection cuts are applied, we fit the invariant mass peaks of the strange particles with a combined gaussian and a linear function to describe the background. Based on the width of the gaussian function, all strange particles inside the $\pm6\sigma$ range around the peak were used to make correlation pairs. We explicitly check that the trigger particle is not a daughter track of an associated V$^{0}$ particle in order to remove any unphysical short-range correlations near the peak region.

The associated per-trigger yield as a function of the azimuthal angle difference $\Delta \varphi$ and pseudorapidity difference $\Delta \eta$ is defined as follows \cite{Abelev:2012ola}:

\begin{equation}
\label{equ:2DCorr1}
\frac{1}{N_{\rm Trig}} \frac{\rm d^{2}\textit{N}_{\rm assoc}}{\rm d(\Delta \varphi)\rm d(\Delta \eta)}  = \frac{1}{N_{\rm Trig}}\frac{N_{\rm sibling}(\Delta \varphi,\Delta \eta)}{N_{\rm mixed}(\Delta \varphi,\Delta \eta)}.
\end{equation}

Here $N_{\rm assoc}$ is the number of associated particles in a given $(\Delta \varphi,\Delta \eta)$ bin and $N_{\rm Trig}$ is the number of trigger particles. The per-trigger yield is measured using the right-hand side of Eq. \ref{equ:2DCorr1}, where $N_{\rm sibling}$ is the total number of pairs in the same event, and $N_{\rm mixed}$ is the number of mixed pairs in mixed events. V$^{0}$s are chosen to have 1 $< p_{\rm T} <$ 6 GeV/\textit{c}. To correct for the geometric acceptance, we used mixed-event distributions where the trigger and associated particles are from two different events. For this purpose, the analysis builds up an event buffer where the tracks of previous events are cached. This buffer stores the events in bins of multiplicity and z-vertex position, and the division is done using events from within these bins to ensure that the events used for the mixed-event distributions have an acceptance that is similar to the events being corrected. The mixed-event distribution is normalized to unity at $(\Delta \varphi,\Delta \eta)=(0,0)$. To increase the statistical significance, we combined the $\Lambda$ and $\bar \Lambda$ signals.

\section{Results}
\begin{figure}[h]
\centering
\begin{tabular}{ll}
\includegraphics[height=2.1in]{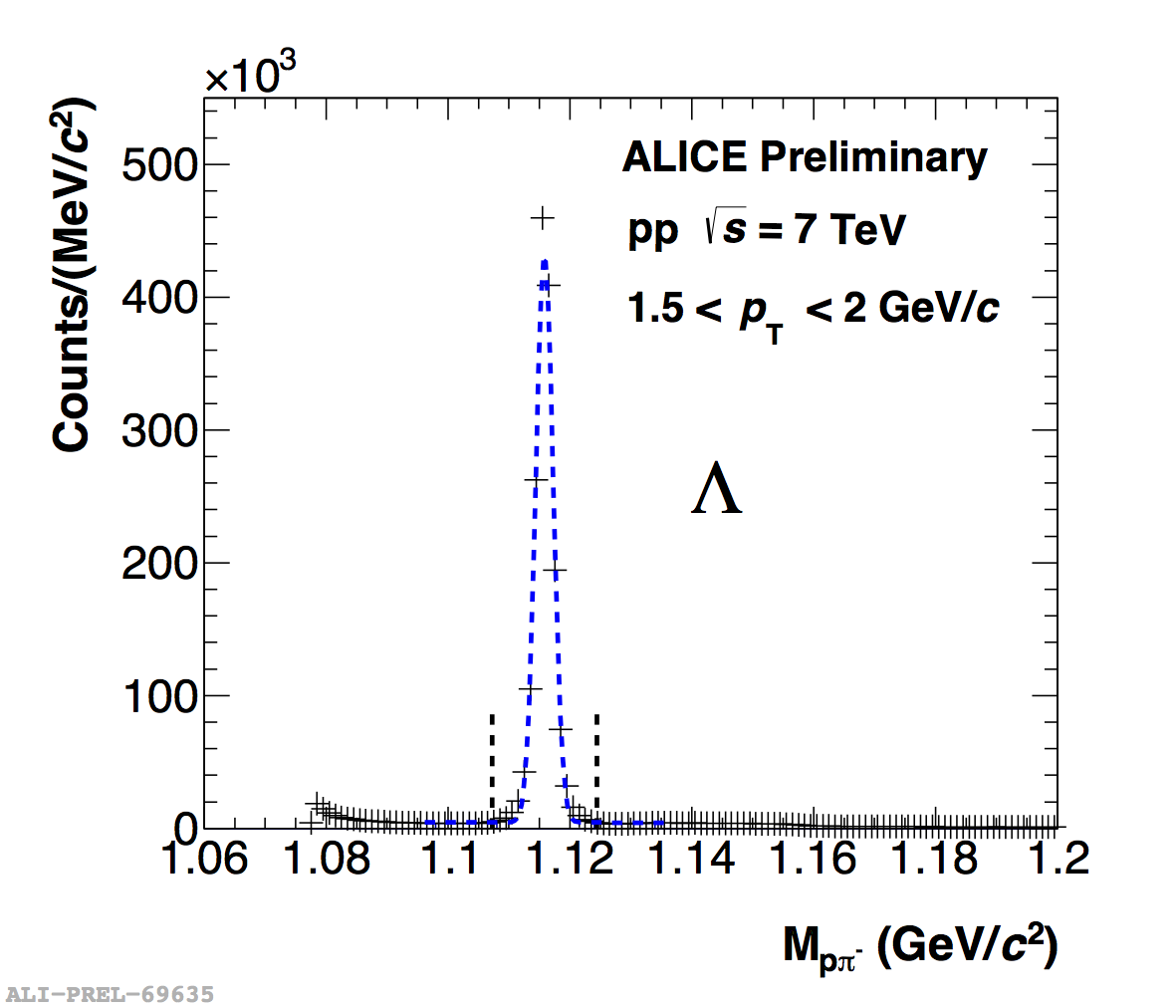}
&
\includegraphics[height=2.1in]{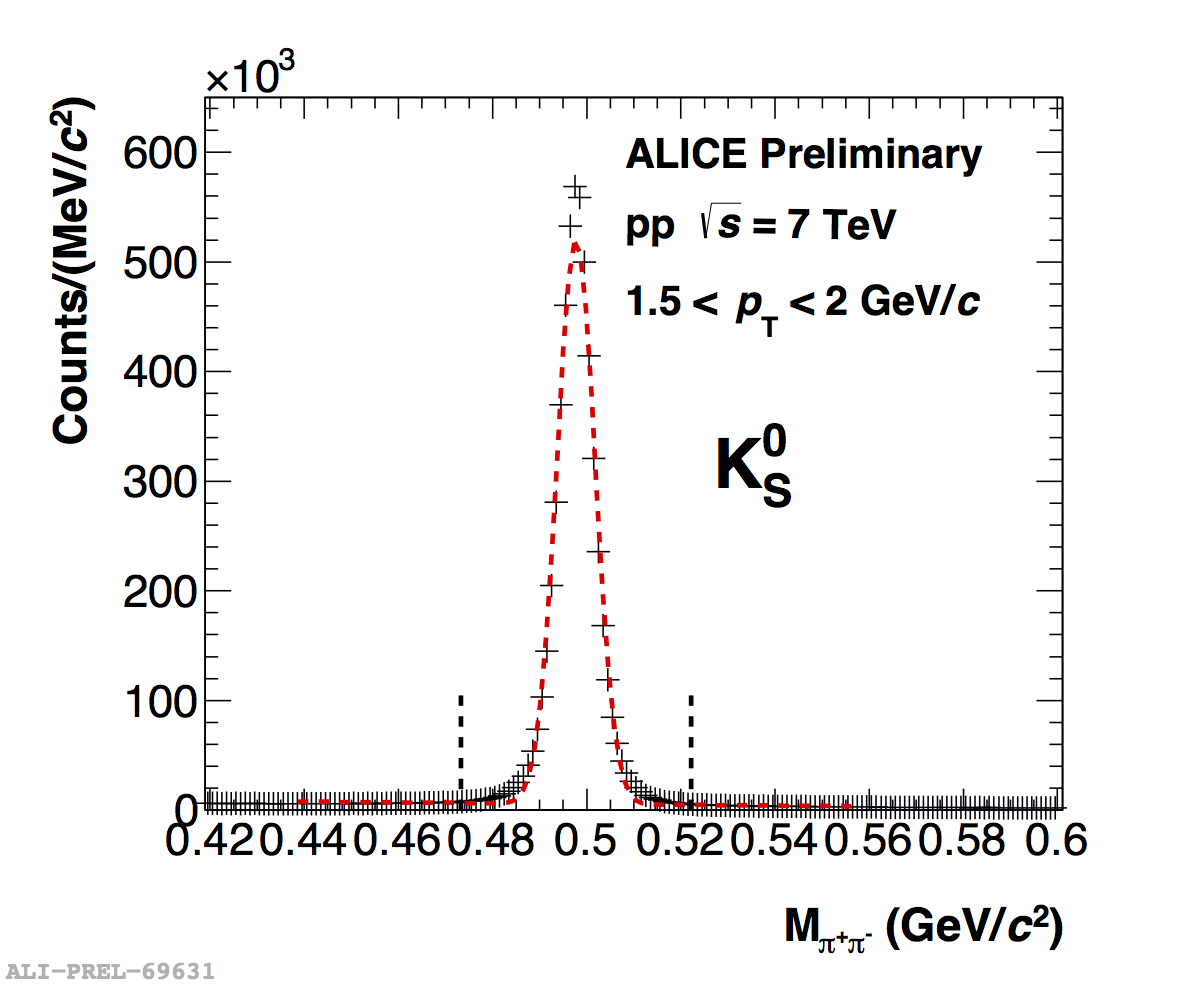}
\end{tabular}
\caption{$\Lambda$(left) and K$^{0}_{\textsubscript{S}}$(right) invariant mass peaks after all selections have been performed. The dashed curve is the fitted Gaussian plus linear function and the two black vertical lines represent the $\pm 6\sigma$ region around the measured mass peak ($\Lambda$ = 1.115 GeV/\textit{c}$^{2}$, K$^{0}_{\textsubscript{S}}$ = 0.497 GeV/\textit{c}$^{2}$).}
\label{fig:figureimass}
\end{figure}

Figure \ref{fig:figureimass} shows the invariant mass peaks for $\Lambda$ and K$^{0}_{\textsubscript{S}}$ after all topological cuts have been applied, fitted with a Gaussian function and a linear function. The $\pm6\sigma$ range used to make correlation pairs with the trigger particle is also shown.
 
\begin{figure}[h]
\centering
\includegraphics[height=2.8in]{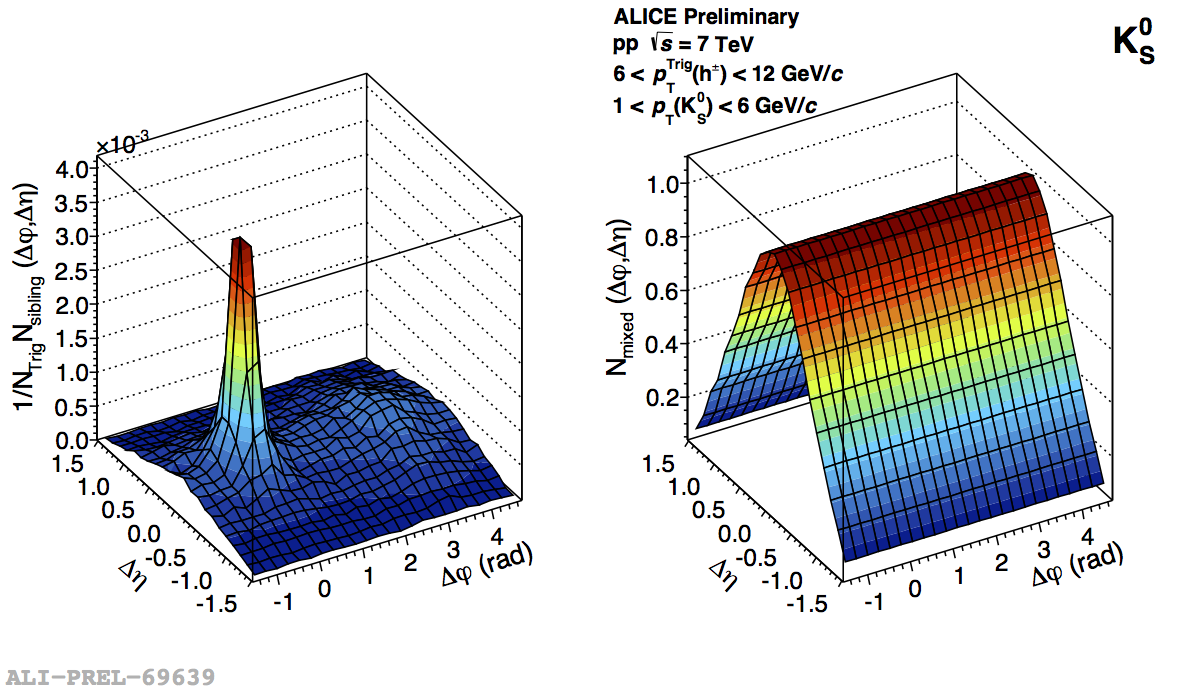}
\caption{$h^{\pm}-$K$_{\textsubscript{S}}^{0}$ correlation for same event (left) and mixed event (right).}
\label{fig:figure1}
\end{figure}

The same-event and mixed-event correlation functions with K$^{0}_{\textsubscript{S}}$ as the associated particles are shown in Figure \ref{fig:figure1}. A prominent near-side peak is visible at $(\Delta \varphi,\Delta \eta)\approx(0,0)$ for the same-event pairs.
 
\begin{figure}[h]
\centering
\includegraphics[height=2.8in]{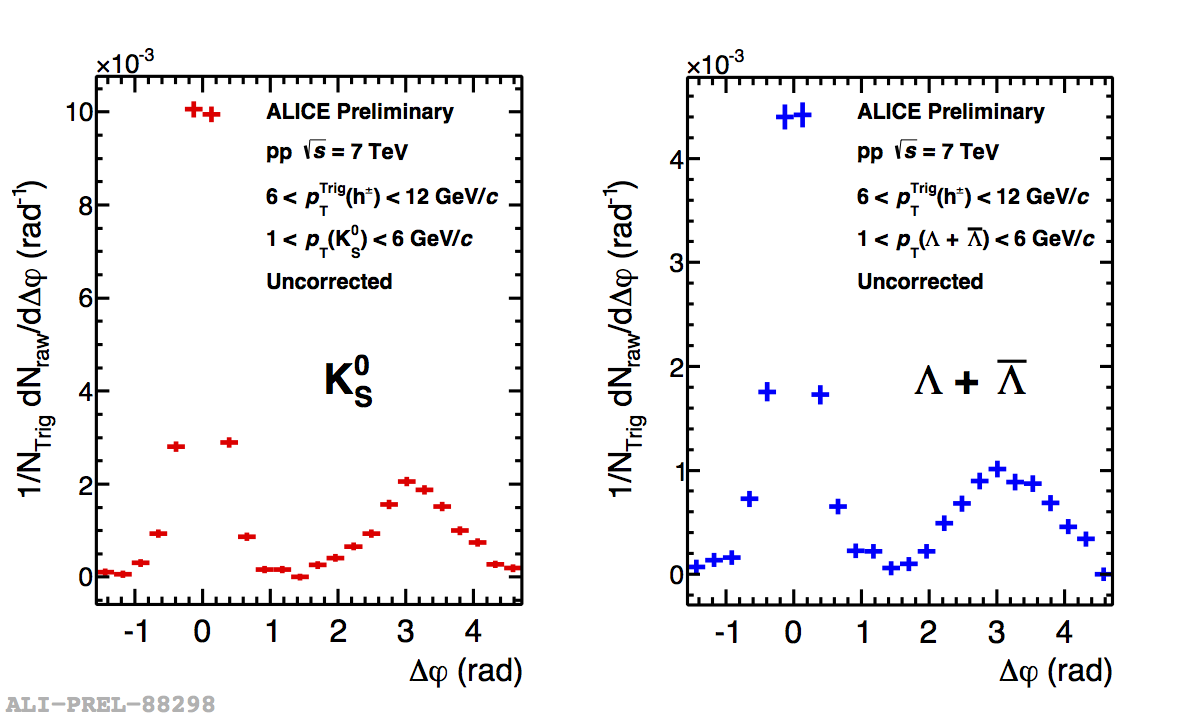}
\caption{Two-particle correlation functions for $h^{\pm}-$K$^{0}_{\textsubscript{S}}$ and $h^{\pm}-(\Lambda+\bar\Lambda)$ pairs at pp $\sqrt{s} = 7$ TeV after the baseline subtraction.}
\label{fig:figure3}
\end{figure}

After the division of the same-event distribution by the mixed-event distribution, we project the 2D correlation function within $|\Delta \eta|<0.5$ to $\Delta \varphi$. To subtract the baseline we find the absolute minimum of the correlation function in the corresponding $p_{\rm T}$ region. Then, we assume that all pairs below this baseline are uncorrelated with the trigger particle and subtract the baseline from the correlation function. This will give us a baseline-subtracted correlation function that has a minimum value of zero.
In Figure \ref{fig:figure3}  we show the projected correlation function in $\Delta \varphi$. In both projections, the near-side jet peak is visible at $\Delta \varphi \approx 0$. 

The smeared peak at $\Delta \varphi \approx \pi$ is from jets produced back-to-back in $\varphi$ and with the parton center-of-mass system (cms) shifted with respect to the cms of the collision. The associated-particle efficiency correction has not yet been applied in these correlation function projections. However, before these can be applied, $\Lambda$ feeddown from the $\Xi^{-}$ and $\Xi^{0}$ decays has to be removed in the near-side, away-side, and underlying-event regions

\begin{figure}[h]
\centering
\begin{tabular}{ll}
\includegraphics[height=2.3in]{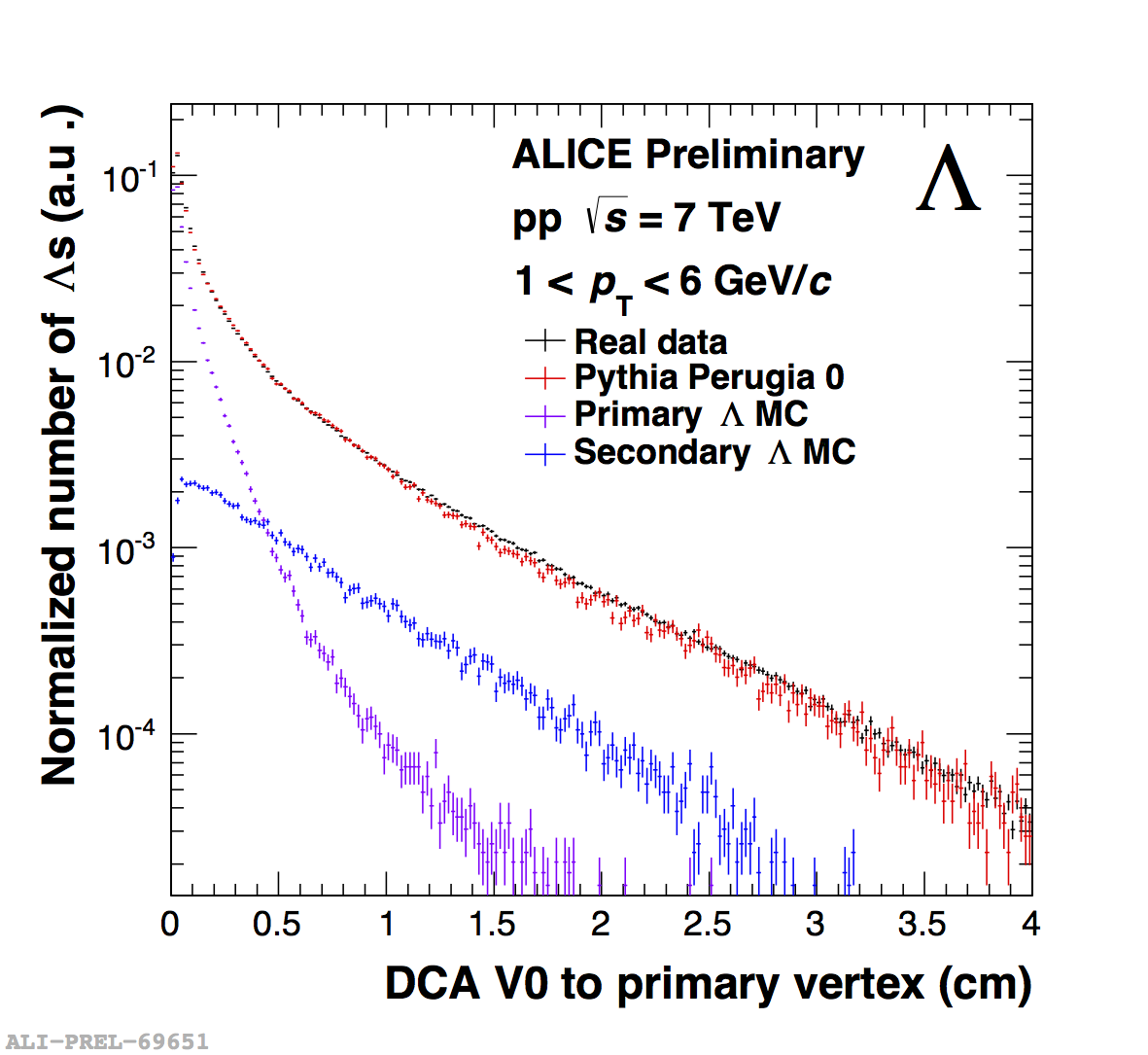}
&
\includegraphics[height=2.3in]{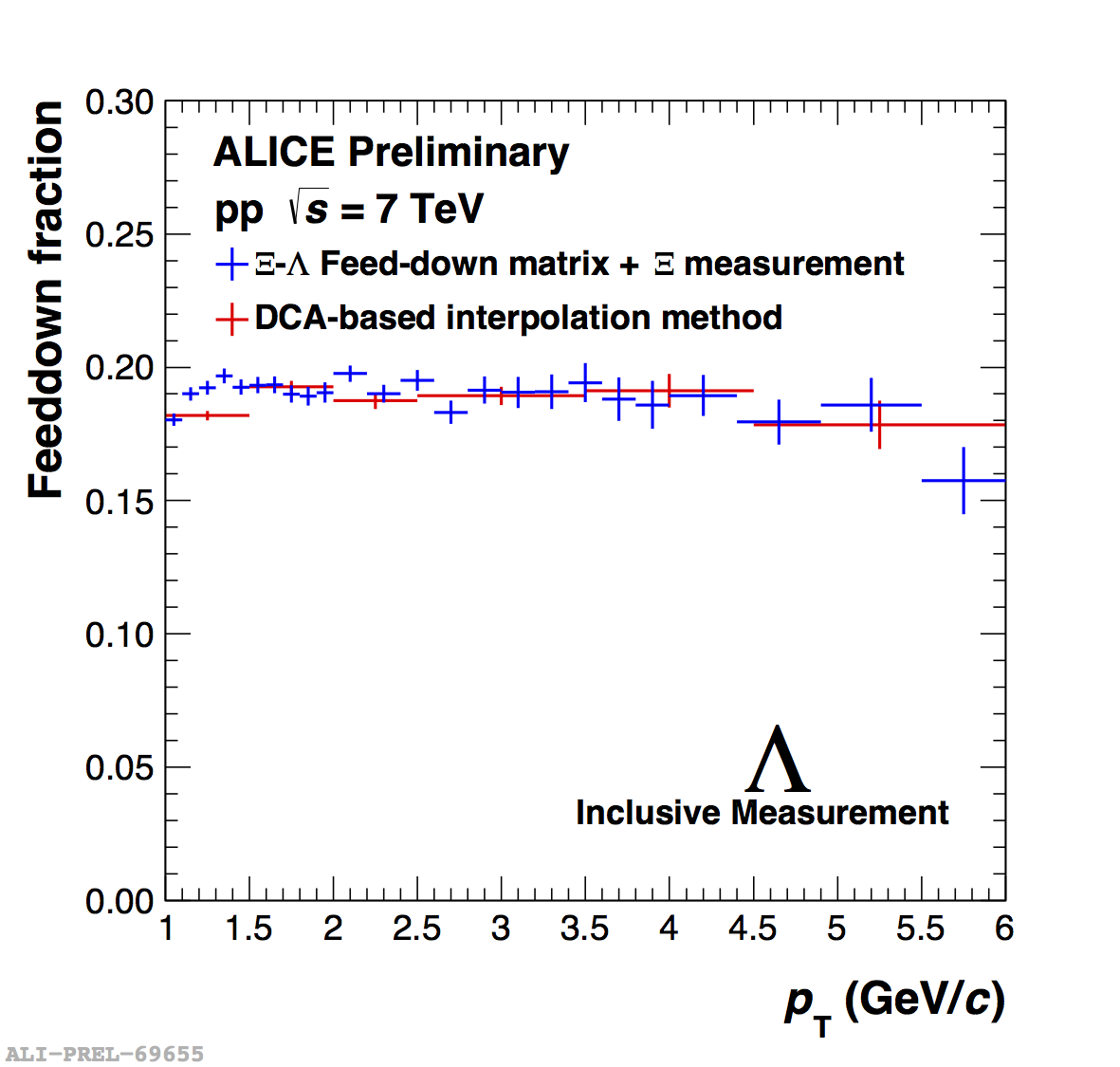}
\end{tabular}
\caption{Estimation of the feeddown fraction of $\Lambda$s from $\Xi^{-}$ and $\Xi^{0}$ using a scaling method based on the DCA to the primary vertex. Left panel: distributions of DCA to the primary vertex of $\Lambda$s in real data (black) and reconstructed MC data (red). The other two distributions are primary (magenta) and secondary (blue) $\Lambda$s, respectively. Only $\Lambda$s from $\Xi^{-}$ and  $\Xi^{0}$ were considered as secondary. Right panel: Feeddown fractions from the more standard $\Xi$ measurement with feeddown matrix (blue) and the scaling method based on the DCA to primary vertex (red).}
\label{fig:figure5}
\end{figure}

The option of using the more traditional inclusive feeddown calculation is not available in this analysis, since no measurement of the  $\Xi^{-}$ and $\Xi^{0}$ production in association with jets and the underlying event is available. \\ 
To overcome this difficulty, we have investigated the use of a data-driven feeddown correction based on the distance of closest approach (DCA) to the primary vertex of secondary and primary V$^{0}$s. The DCA distributions for these two contributions are different enough that a change of DCA selection in the analysis will result in the removal of a different fraction of each contribution.  When varying the DCA selection, one can use the relative signal change observed in data to determine the fraction of secondaries present in the analysis via:

\begin{equation}
\label{equ:DCAscale}
\left(\frac{\Delta S^{Data}}{S^{Data}}\right) = f_{prim}\left(\frac{\Delta S^{MC}_{prim}}{S^{MC}_{prim}}\right)+f_{sec}\left(\frac{\Delta S^{MC}_{sec}}{S_{sec}}\right) = \left(\frac{\Delta S^{MC}_{prim}}{S^{MC}_{prim}}\right)+f_{sec}\left[\left(\frac{\Delta S^{MC}_{sec}}{S_{sec}}\right)-\left(\frac{\Delta S^{MC}_{prim}}{S^{MC}_{prim}}\right)\right],
\end{equation}
where $S^{Data}$ is the total signal observed in data, $S^{MC}_{prim/sec}$ are the primary and secondary signals as seen in simulations, $\Delta{S}$ denotes, in all cases, the observed signal variation when changing the DCA selection, and $f_{prim/sec}$ is the fraction of primary and secondary V$^{0}$s in the real data, where  $f_{prim}$+ $f_{sec}$=1.  As a benchmark, the inclusive\footnote{Total $\Lambda$ yields from the near side, away side and underlying event.} $f_{sec}$ computed with this method compares favorably with the fraction of secondaries computed with a more traditional approach that utilizes a feeddown matrix
and a $\Xi$ measurement, as can be seen in Figure \ref{fig:figure5}.  This result is encouraging, since the DCA-based interpolation method has the advantage that one can determine a feeddown fraction in each region (near and away side, underlying event) separately. Work is currently ongoing to finalize these yields, which includes incorporating the associated particle efficiency corrections as well as performing track splitting studies.

\section{Conclusions}
We have shown the uncorrected two-particle correlation function for $h^{\pm}-$K$^{0}_{\textsubscript{S}}$ and $h^{\pm}-(\Lambda+\bar\Lambda)$. The near-side and away-side regions, which originate from parton fragmentation, are clearly visible even without full jet reconstruction, demonstrating the potential of using azimuthal correlations to probe jet-sensitive physics. Further, we estimate the feeddown fraction of $\Lambda$s from $\Xi^{-}$ and $\Xi^{0}$ using a scaling method based on the DCA of the V$^{0}$s to the primary vertex. Results were shown to be comparable with the ones obtained with a feeddown correction approach that utilizes a $\Xi$ measurement, demonstrating that there is no need for a $\Xi$ measurement in order to compute primary $\Lambda$ yields.
\Acknowledgements
The work was supported by the grant DE-FG02-07ER41521 funded by the U.S. Department of Energy, Office of Science.

\end{document}